\documentstyle[aps,pra,epsfig]{revtex}

\begin{document}

\draft

\title{Continuous variable teleportation 
of a quantum state onto a macroscopic body}

\author{
Stefano Mancini,
David Vitali,
and Paolo Tombesi}

\address{
INFM, Dipartimento di Fisica,
Universit\`a di Camerino,
I-62032 Camerino, Italy}

\date{\today}

\maketitle

\begin{abstract}
We study the possibility to teleport an unkown quantum state 
onto the vibrational degree of freedom of a movable mirror.
The quantum channel between the two parties 
is established by exploiting radiation pressure effects.
\end{abstract}


\section{Introduction}

Quantum state teleportation is undoubtedly one of the most fascinating
developments of quantum information processing \cite{BEN93}.

Teleportation of an unknown quantum state is its immaterial
transport through a classical channel employing one of the 
most puzzling resources of Quantum
Mechanics: entanglement
\cite{EIN35}.
A variety of possible experimental schemes have been proposed and few of
them partially realized in the discrete variable case
involving the polarization state of single photons\cite{BOU97,BOS98,JEN02}.
A successful achievement has been then obtained in the continuous variable case
of an optical field \cite{FUR98}.
However, the tantalizing problem of
extending quantum teleportation at the macroscopic scale
still remains open.

Recently, in the perspective of demonstrating and manipulating
the quantum properties of bigger and bigger objects \cite{JUL01},
it has been shown \cite{PRL02} how it is possible
to entangle two massive macroscopic oscillators,
like movable mirrors, by using radiation pressure effects.
The creation of such an entanglement at the macroscopic level suggests
an avenue for achieving teleportation of a continuous variable state
of a radiation field onto the vibrational state of a mirror.

\section{The Model}

\begin{figure}[t]
\centerline{\epsfig{figure=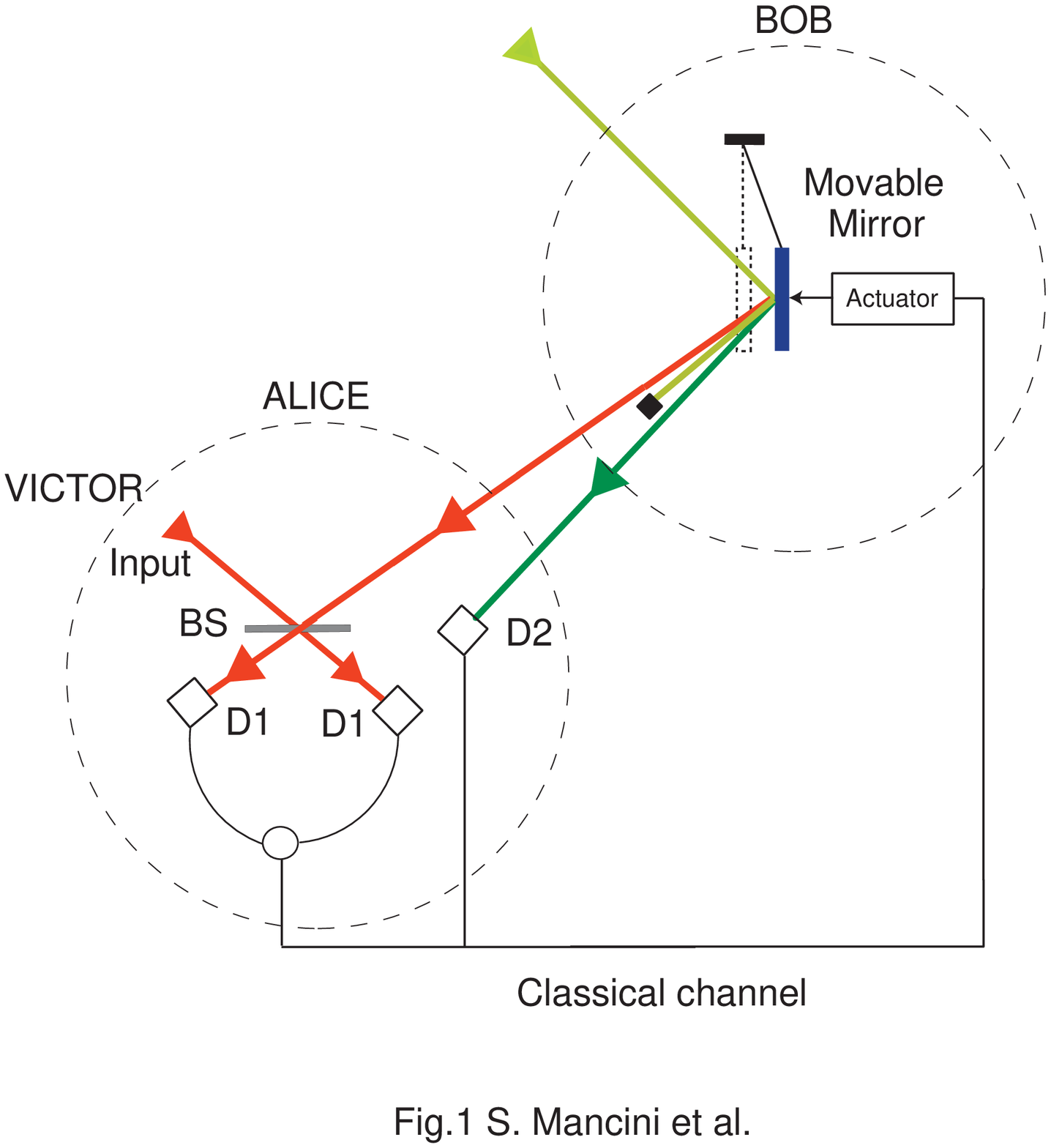,width=3.0in}}
\caption{\widetext
Schematic description of the system. A laser field at frequency
$\omega_{0}$ impinges on the mirror oscillating at frequency $\Omega$.
In the reflected field two sideband modes are excited at
frequencies $\omega_{1}=\omega_{0}-\Omega$ and
$\omega_{2}=\omega_{0}+\Omega$.
These two modes then reach Alice's station.
The mode at frequency $\omega_{2}$ is subjected to a heterodyne
measurement $D2$, while the mode at frequency $\omega_{1}$
is mixed in the 50-50 beam splitter BS with the unknown input given by Victor.
A Bell-like measurement $D1$ is then performed
on this combination and the result, combined with the
heterodyne one, is fed-forward to Bob
as two bits of classical information.
Finally, he actuates the displacement
in the phase space of the moving mirror.
}
\label{fig1}
\end{figure}

We consider the situation where an unknown
quantum state of a radiation field is prepared by a verifier
(Victor) and sent to an analyzing station (Alice).
Here we shall provide a protocol
which enables Alice to teleport the unknown quantum state of the radiation
onto a collective vibrational degree of freedom
of a macroscopic, perfectly reflecting, mirror placed at a remote station
(Bob) (see Fig.~\ref{fig1}).
For simplicity we consider only the motion and the elastic deformations
of the mirror taking place along the spatial direction $x$,
orthogonal to its reflecting surface. 
Then we consider an intense laser beam
impinging on the surface of the mirror, 
whose radiation pressure realizes an optomechanical coupling \cite{SAM95}.
In fact, the electromagnetic field exerts a force on the mirror 
proportional to its intensity and, at the same time,
it is phase-shifted by
the mirror displacement from the equilibrium position \cite{LAW95}.
In the limit of small mirror displacements, and in the interaction
picture with respect to the free Hamiltonian of the electromagnetic field
and the mirror displacement field ${\hat x({\bf r},t)}$
(${\bf r}$ is the coordinate on the mirror surface), one has the
following Hamiltonian
 \cite{PIN99}
\begin{eqnarray}
    {\hat H}&=&-\int\,d^{2}{\bf r}\,
    {\hat P}({\bf r},t){\hat x}({\bf r},t)\,,
    \label{eq:Hini}
\end{eqnarray}
where ${\hat P({\bf r},t)}$ is the radiation pressure force \cite{SAM95}.
All the continuum of electromagnetic modes
with positive longitudinal wave vector $q$, transverse
wave vector ${\bf k}$, and frequency $\omega=\sqrt{c^{2}(k^{2}+q^{2})}$
($c$ being the light speed in the vacuum)
contributes to the radiation pressure force.
We are adopting the interaction 
picture with respect to the free Hamiltonian of the electromagnetic 
field of the continuum and of the field of elastic deformations of 
the mirror.
Following Ref.\cite{SAM95}, and considering linearly polarized radiation
with the electric field parallel to the mirror surface, we have 
\begin{eqnarray}
        {\hat P}({\bf r},t)&=&-\frac{\hbar}{8\pi^{3}}
        \int d{\bf k}\int dq\int d{\bf k'}\int dq'
        \frac{c^{2}qq'}{\sqrt{\omega\omega'}} 
        ({\bf u}_{k}\cdot{\bf u}_{k'}){\bf u}_{q}
        \nonumber\\
        &\times&
        \left\{{\hat a}({\bf k},q){\hat a}({\bf k}',q')
        \exp[-i(\omega+\omega')t
        +i({\bf k}+{\bf k}')\cdot{\bf r}]\right.
        \nonumber\\
        &&\left.
        +{\hat a}^{\dag}({\bf k},q){\hat a}^{\dag}({\bf k}',q')
        \exp[i(\omega+\omega')t
        -i({\bf k}+{\bf k}')\cdot{\bf r}]\right.
        \nonumber\\
        &&\left. 
        +{\hat a}({\bf k},q){\hat a}^{\dag}({\bf k}',q')
        \exp[-i(\omega-\omega')t
        +i({\bf k}-{\bf k}')\cdot{\bf r}]\right.
        \nonumber\\
        &&\left.
        +{\hat a}^{\dag}({\bf k},q){\hat a}({\bf k}',q')
        \exp[i(\omega-\omega')t
        -i({\bf k}-{\bf k}')\cdot{\bf r}]
        \right\}\,,
        \label{eq:P}
\end{eqnarray}
where ${\hat a}({\bf k},q)$ are the continuous mode destruction 
operators having transverse wave vector ${\bf k}$ and positive 
longitudinal wave vector component $q$, obeying the commutation 
relations
\begin{equation}
\left[{\hat a}({\bf k},q),{\hat a}({\bf k}',q')\right]
=\delta({\bf k}-{\bf k}')\delta(q-q')\,.
\end{equation}
Furthermore, the electromagnetic wave frequencies 
$\omega$ and $\omega'$ are given by $\omega^{2}=c^{2}(k^{2}+q^{2})$
and $\omega'^{2}=c^{2}(k'^{2}+q'^{2})$, 
and ${\bf u}_{k}$, ${\bf u}_{q}$ denote dimensionless
unit vectors parallel to ${\bf k}$, $q$ respectively.

The mirror displacement ${\hat x({\bf r},t)}$ is generally given by a 
superposition of many acoustic modes \cite{PIN99};
however, a single vibrational mode description can be adopted whenever 
detection is limited to a frequency bandwidth
including a single mechanical resonance. 
In particular, focused light beams are able to excite 
Gaussian acoustic modes, in which only a small portion of the mirror,
localized at its center, vibrates. These modes have a small  
waist $w$, a large mechanical quality 
factor $Q$, a small effective mass $M$ \cite{PIN99}, and 
the simplest choice is to choose the fundamental Gaussian mode with 
frequency $\Omega$, i.e., 
\begin{equation}
    {\hat x}({\bf r},t)=\sqrt{\frac{\hbar}{2M\Omega}}
    \left[{\hat b} e^{-i\Omega t}+{\hat b}^{\dag}e^{i\Omega t}\right]
    \exp(-r^{2}/w^{2})\,.
    \label{eq:x}
\end{equation}
By inserting Eqs.(\ref{eq:P}) and (\ref{eq:x}) in Eq.(\ref{eq:Hini})
and integrating over the variable ${\bf r}$ one obtains 
\begin{eqnarray}
        {\hat H}&=&-\frac{\hbar 
        w^{2}}{8\pi^{2}}\sqrt{\frac{\hbar}{2M\Omega}}
        \int d{\bf k}\int dq\int d{\bf k'}\int dq'
        \frac{c^{2}qq'}{\sqrt{\omega\omega'}} 
        ({\bf u}_{k}\cdot{\bf u}_{k'}){\bf u}_{q}
        \nonumber\\
        &\times&
        \left\{{\hat a}({\bf k},q){\hat a}({\bf k}',q')
        \exp[-i(\omega+\omega')t
        -({\bf k}+{\bf k}')^{2}w^{2}/4]\right.
        \nonumber\\
        &&\left.
        +{\hat a}^{\dag}({\bf k},q){\hat a}^{\dag}({\bf k}',q')
        \exp[i(\omega+\omega')t
        -({\bf k}+{\bf k}')^{2}w^{2}/4]\right.
        \nonumber\\
        &&\left. 
        +{\hat a}({\bf k},q){\hat a}^{\dag}({\bf k}',q')
        \exp[-i(\omega-\omega')t
        -({\bf k}-{\bf k}')^{2}w^{2}/4]\right.
        \nonumber\\
        &&\left.
        +{\hat a}^{\dag}({\bf k},q){\hat a}({\bf k}',q')
        \exp[i(\omega-\omega')t
        -({\bf k}-{\bf k}')^{2}w^{2}/4]
        \right\}\times\left\{{\hat b}e^{-i\Omega t}
        +{\hat b}^{\dag}e^{i\Omega t}\right\}\,.
        \label{eq:Hint1}
\end{eqnarray}
In common situations, the acoustical waist $w$ is much larger than typical 
optical wavelengths \cite{PIN99}, and therefore we can approximate
$\exp\left\{-({\bf k}\pm {\bf k'})^{2}w^{2}/4\right\}w^{2}/4\pi \simeq 
\delta({\bf k}\pm {\bf k'})$ and then integrate Eq.~(\ref{eq:Hint1})
over ${\bf k'}$, obtaining
\begin{eqnarray}
        {\hat H}&=&-\frac{\hbar}{2\pi}\sqrt{\frac{\hbar}{2M\Omega}}
        \int d{\bf k}\int dq\int dq'
        \frac{c^{2}qq'}{\sqrt{\omega\omega'}} 
        \nonumber\\
        &\times&
        \left\{-{\hat a}({\bf k},q){\hat a}(-{\bf 
        k},q')\exp[-i(\omega+\omega')t]
        \right.
        \nonumber\\
        &&\left.
        -{\hat a}^{\dag}({\bf k},q){\hat a}^{\dag}(-{\bf 
        k},q')\exp[i(\omega+\omega')t]\right.
        \nonumber\\
        &&\left. 
        +{\hat a}({\bf k},q){\hat a}^{\dag}({\bf 
        k},q')\exp[-i(\omega-\omega')t]\right.
        \nonumber\\
        &&\left.
        +{\hat a}^{\dag}({\bf k},q){\hat a}({\bf k},q')
        \exp[i(\omega-\omega')t]
        \right\}\times\left\{{\hat b}e^{-i\Omega t}+{\hat 
        b}^{\dag}e^{i\Omega t}\right\}\,.
        \label{eq:Hint2}
\end{eqnarray}
We now make the Rotating Wave Approximation (RWA), that is, we 
neglect all the terms oscillating in time faster than the mechanical 
frequency $\Omega$. This means averaging the Hamiltonian over a time 
$\tau$ such 
that $\Omega \tau \gg 1$, yielding the following replacements in 
Eq.~(\ref{eq:Hint2})
\begin{equation}
\exp\left\{\pm i(\omega' \pm \omega \pm \Omega)t\right\} \rightarrow 
\frac{2\pi}{\tau}\delta(\omega' \pm \omega \pm \Omega).
\end{equation}
The parameter $\tau$ is not arbitrary, but its inverse, $1/\tau = 
\Delta\nu_{det}$, is the 
detection bandwidth, that is, the spectral resolution of the 
detection apparata used at Alice station. 

Since $\omega$ and $\omega'$ are positive and $\Omega$ is much 
smaller than typical optical frequencies, the two terms 
$\delta(\omega' + \omega \pm \Omega)$ give no contribution, while the 
other two terms can be rewritten as
\begin{equation}
\frac{2\pi}{\tau}\delta(\omega' - \omega \pm \Omega) =
2\pi\Delta\nu_{det}\delta(q' - 
\bar{q}_{\pm})\frac{\omega'(\bar{q}_{\pm})}{c^{2}\bar{q}_{\pm}},
\end{equation}
where $\bar{q}_{\pm}=\sqrt{(\omega \pm \Omega)^{2}/c^{2}-k^{2}}$.
Integrating over $q'$ we get
\begin{eqnarray}
        {\hat H}&=&-\hbar\Delta\nu_{det}\sqrt{\frac{\hbar}{2M\Omega}}
        \int d{\bf k}\int dq
        \frac{q}{\sqrt{\omega}} 
        \left\{{\hat a}({\bf k},q){\hat a}^{\dag}\left({\bf 
        k},\bar{q}_{+}\right){\hat b}\;
        \sqrt{\omega+\Omega}
        +{\hat a}({\bf k},q){\hat a}^{\dag}\left({\bf k},
        \bar{q}_{-}\right){\hat b}^{\dag}\sqrt{\omega-\Omega}
        \right.
        \nonumber\\
        &&\left. 
        +{\hat a}^{\dag}({\bf k},q){\hat a}\left({\bf k},
        \bar{q}_{+}\right){\hat b}^{\dag}
        \sqrt{\omega+\Omega}
        +{\hat a}^{\dag}({\bf k},q){\hat a}\left({\bf k},
        \bar{q}_{-}\right){\hat b}\sqrt{\omega-\Omega}
        \right\}\,,
        \label{eq:Hint3}
\end{eqnarray}
where we have used the fact that $\omega'(\bar{q}_{\pm})=\omega \pm 
\Omega$.

We now consider the situation where the radiation field incident on 
the mirror is characterized by an intense, quasi-monochromatic, 
laser field with trasversal 
wave vector ${\bf k_{0}}$, longitudinal wave vector $q_{0}$, 
cross-sectional area $A$, and power ${\wp}$. Since this component is 
very intense, it can be 
treated as classical and one can approximate
${\hat a}({\bf k},q) \simeq  \alpha({\bf k},q)$ in Eq.~(\ref{eq:Hint3}), 
where (with an appropriate choice of phases)
\begin{equation}
\alpha({\bf k},q) = -i\sqrt{\frac{(2\pi)^{3}{\wp}}{\hbar \omega_{0}cA}}
\delta({\bf k}-{\bf k_{0}})\delta(q-q_{0})\,,
\label{intenso}
\end{equation}
with $\omega_{0}=c\sqrt{{\bf k_{0}}^{2}+q_{0}^{2}}$.

Due to the Dirac delta, the only nonvanishing terms in the 
optomechanical interaction driven by the intense laser beam
involve only two back-scattered waves, that is, the sidebands of the driving 
beam at frequencies 
$\omega_{0}\pm \Omega$, as described by 
\begin{eqnarray}
        {\hat H}&=& i\hbar\Delta\nu_{det}\sqrt{\frac{\hbar}{2M\Omega}}
        q_{0} \sqrt{\frac{{\wp}}{\hbar \omega_{0}cA}}
        \left\{\sqrt{\frac{\omega_{0}+\Omega}{\omega_{0}}}
        {\hat a}^{\dag}\left({\bf 
        k_{0}},\bar{q}_{+}\right){\hat b}
        +\sqrt{\frac{\omega_{0}-\Omega}{\omega_{0}}}
        {\hat a}^{\dag}\left({\bf 
        k_{0}},\bar{q}_{-}\right){\hat b}^{\dagger}
        \right.
        \nonumber\\
        &&\left. 
        -\sqrt{\frac{\omega_{0}+\Omega}{\omega_{0}}}
        {\hat a}\left({\bf 
        k_{0}},\bar{q}_{+}\right){\hat b}^{\dagger}
        -\sqrt{\frac{\omega_{0}-\Omega}{\omega_{0}}}
        {\hat a}\left({\bf 
        k_{0}},\bar{q}_{-}\right){\hat b}\right\},
        \label{eq:Heff0}
\end{eqnarray}
where now $\bar{q}_{\pm}=\sqrt{(\omega_{0} \pm \Omega)^{2}/c^{2}-k_{0}^{2}}$.
The physical process described by this interation Hamiltonian is 
very similar to a stimulated Brillouin scattering \cite{PER84}, even though in 
this case the Stokes and anti-Stokes component are back-scattered by 
the acoustic waves at 
reflection, and the optomechanical coupling is provided by the 
radiation pressure
and not by the dielectric properties of the mirror.

In practice, either the driving laser beam and the back-scattered modes
are never monochromatic, but have a nonzero bandwidth. In general the 
bandwidth of the back-scattered modes is determined by the bandwidth 
of the driving laser beam and that of the acoustic mode. However, due 
to its high mechanical quality factor, the spectral width of the 
mechanical resonance is negligible (about $1$ Hz) and, in practice, the 
bandwidth of the two sideband modes $\Delta \nu_{mode}$
coincides with that of the incident laser beam.
It is then convenient to consider this nonzero bandwidth to redefine
the bosonic operators of the Stokes and anti-Stokes modes 
to make them dimensionless, 
\begin{eqnarray}
{\hat a}_{1}&=& 2\pi \sqrt{\frac{2\pi \Delta \nu_{mode}}{cA}}
{\hat a}\left({\bf k_{0}},\bar{q}_{-}\right) =
2 \pi \sqrt{\frac{\Delta q}{A}}{\hat a}\left({\bf k_{0}},\bar{q}_{-}\right)\\
{\hat a}_{2}&=& 2\pi \sqrt{\frac{2\pi \Delta \nu_{mode}}{cA}}
{\hat a}\left({\bf k_{0}},\bar{q}_{+}\right)=
2 \pi \sqrt{\frac{\Delta q}{A}}{\hat a}\left({\bf k_{0}},\bar{q}_{+}\right), 
\end{eqnarray}
so that  Eq.(\ref{eq:Heff0}) reduces to an effective 
Hamiltonian
\begin{equation}
    {\hat H}_{eff}=-i\hbar \chi
    ({\hat a}_{1}{\hat b}-{\hat a}^{\dag}_{1}{\hat b}^{\dag})
    -i\hbar\theta({\hat a}_{2}{\hat b}^{\dag}-{\hat a}^{\dag}_{2}
    {\hat b})\,,
    \label{eq:Heff}
\end{equation}
where the couplings $\chi$ and $\theta$ are given by
\begin{eqnarray}
\label{chi}
\chi &=& q_{0}\Delta\nu_{det}\sqrt{\frac{\hbar}{2M\Omega}}
\sqrt{\frac{{\wp} }{ \Delta \nu_{mode}\hbar \omega_{0}}}
\sqrt{\frac{\omega_{0}-\Omega}{\omega_{0}}}=
\cos\phi_{0}\sqrt{\frac{{\wp}\Delta\nu_{det}^{2}(\omega_{0}-\Omega)}{2M\Omega 
c^{2}\Delta\nu_{mode}}} \\
\theta &=& \chi \sqrt{\frac{\omega_{0}+\Omega}{\omega_{0}-\Omega}},
\end{eqnarray}
with $\phi_{0}=\arccos(cq_{0}/\omega_{0})$, 
is the angle of incidence of the driving beam. 
It is possible to verify that with the above definitions, the Stokes 
and anti-Stokes annihilation operators $a_{1}$ and $a_{2}$ satisfy the 
usual commutation relations 
$\left[a_{i},a_{j}^{\dagger}\right]=\delta_{i,j}$.

\section{System dynamics}

Eq.~(\ref{eq:Heff}) contains two interaction terms: the first one,
between modes ${\hat a}_{1}$ and ${\hat b}$,
is a parametric-type interaction
leading to squeezing in phase space \cite{QO94}, and it is
able to generate the EPR-like
entangled state which has been used in the continuous variable teleportation
experiment of Ref.~\cite{FUR98}. The
second interaction term, between modes ${\hat a}_{2}$ and ${\hat b}$,
is a beam-splitter-type
interaction \cite{QO94}, which may degrade the entanglement between
modes ${\hat a}_{1}$ and ${\hat b}$ generated by the first term.

The Hamiltonian (\ref{eq:Heff}) leads to a system of 
linear Heisenberg equations, namely 
\begin{mathletters}
\begin{eqnarray}
    \dot{\hat a}_{1}&=&\chi {\hat b}^{\dag}\,,
    \\
    \dot{\hat b}&=&\chi {\hat a}_{1}^{\dag}-\theta {\hat a}_{2}\,,
    \\
    \dot {\hat a}_{2}&=&\theta {\hat b}\,.
\end{eqnarray}
\end{mathletters}
The solutions read
\begin{mathletters}\label{eq:sol}
\begin{eqnarray}
    {\hat a}_{1}(t)&=&\frac{1}{\Theta^{2}}\left[\theta^{2}-\chi^{2}
    \cos\left(\Theta t\right)\right]{\hat a}_{1}(0)
    +\frac{\chi}{\Theta}\sin\left(\Theta t\right) {\hat b}^{\dag}(0)
    -\frac{1}{\Theta^{2}}\left[\chi\theta-\chi\theta
    \cos\left(\Theta t\right)\right]{\hat a}_{2}^{\dag}(0)\,,
    \\
    {\hat b}(t)&=&-\frac{\chi}{\Theta}\sin\left(\Theta t\right) 
    {\hat a}_{1}^{\dag}(0)
    +\cos\left(\Theta t\right) {\hat b}(0)
    -\frac{\theta}{\Theta}\sin\left(\Theta t\right) {\hat a}_{2}(0)\,,
    \\
    {\hat a}_{2}(t)&=&\frac{1}{\Theta^{2}}\left[\chi\theta-\chi\theta
    \cos\left(\Theta t\right)\right]{\hat a}_{1}^{\dag}(0)
    -\frac{\theta}{\Theta}\sin\left(\Theta t\right) {\hat b}(0)
    -\frac{1}{\Theta^{2}}\left[\chi^{2}-\theta^{2}
    \cos\left(\Theta t\right)\right]{\hat a}_{2}(0)\,,
\end{eqnarray}
\end{mathletters}
where $\Theta=\sqrt{\theta^{2}-\chi^{2}}$.

On the other hand, the system dynamics can be easily studied also through
the (normally ordered) characteristic function $\Phi(\mu,\nu,\zeta)$,
where $\mu,\nu,\zeta$ are the complex variables corresponding
to the operators ${\hat a}_{1},{\hat b},{\hat a}_{2}$ respectively.
From the Hamiltonian (\ref{eq:Heff}) the dynamical equation for 
$\Phi$ results
\begin{eqnarray}\label{eq:Phidot}
    {\dot\Phi}&=&\chi\left(
    \mu\nu+\mu^{*}\nu^{*}-\mu^{*}\frac{\partial}{\partial\nu}
    -\mu\frac{\partial}{\partial\nu^{*}}
    -\nu^{*}\frac{\partial}{\partial\mu}
    -\nu\frac{\partial}{\partial\mu^{*}}\right)\Phi
    \nonumber\\
    &&+\theta\left(
    \zeta^{*}\frac{\partial}{\partial\nu^{*}}
    +\zeta\frac{\partial}{\partial\nu}
    -\nu^{*}\frac{\partial}{\partial\zeta^{*}}
    -\nu\frac{\partial}{\partial\zeta}\right)\Phi\,,
\end{eqnarray}
with the initial condition
\begin{equation}\label{eq:Phiini}
    \Phi(t=0)=\exp\left[-\overline{n}|\nu|^{2}\right]\,,
\end{equation}
corresponding to the vacuum for the modes ${\hat a}_{1}$,
${\hat a}_{2}$ and to a thermal state for the mode ${\hat b}$.
The latter is characterized by an average number of excitations
$\overline{n}=[\coth(\hbar\Omega/2k_{B}T)-1]/2$,
$T$ being the equilibrium temperature and $k_{B}$ the
Boltzmann constant.
Then, equation (\ref{eq:Phidot}) has a Gaussian solution of the form
\begin{equation}\label{eq:Phisol}
    \Phi=\exp\left[
    -{\cal A}|\mu|^{2}-{\cal B}|\nu|^{2}-{\cal E}|\zeta|^{2}
    +{\cal C}\mu\nu+{\cal C}\mu^{*}\nu^{*}
    +{\cal F}\mu\zeta+{\cal F}\mu^{*}\zeta^{*}
    +{\cal D}\nu\zeta^{*}+{\cal D}\nu^{*}\zeta\right]\,,
\end{equation}
where
\begin{mathletters}\label{eq:ABCDEF}
\begin{eqnarray}
    {\cal A}(t)&=&\frac{\chi^{4}}{2\Theta^{4}}
    \left[\cos\left(2\Theta t\right)-1\right]
    -2\frac{\chi^{2}\theta^{2}}{\Theta^{4}}
    \left[\cos\left(\Theta t\right)-1\right]
    +\overline{n}\frac{\chi^{2}}{2\Theta^{2}}
    \left[1-\cos\left(2\Theta t\right)\right]\,,
    \\
    {\cal B}(t)&=&\frac{\chi^{2}}{2\Theta^{2}}
    \left[1-\cos\left(2\Theta t\right)\right]
    +\frac{\overline{n}}{2}
    \left[1+\cos\left(2\Theta t\right)\right]\,,
    \\
    {\cal C}(t)&=&-\frac{\chi^{3}}{2\Theta^{3}}
    \sin\left(2\Theta t\right)
    +\frac{\chi\theta^{2}}{\Theta^{3}}
    \sin\left(\Theta t\right)
    +\overline{n}\frac{\chi}{2\Theta}
    \sin\left(2\Theta t\right)\,,
    \\
    {\cal D}(t)&=&\frac{\chi^{2}\theta}{2\Theta^{3}}
    \sin\left(2\Theta t\right)
    -\frac{\chi^{2}\theta}{\Theta^{3}}
    \sin\left(\Theta t\right)
    -\overline{n}\frac{\theta}{2\Theta}
    \sin\left(2\Theta t\right)\,,
    \\
    {\cal E}(t)&=&\frac{\chi^{2}\theta^{2}}{2\Theta^{4}}
    \left[\cos\left(2\Theta t\right)-1\right]
    -2\frac{\chi^{2}\theta^{2}}{\Theta^{4}}
    \left[\cos\left(\Theta t\right)-1\right]
    +\overline{n}\frac{\theta^{2}}{2\Theta^{2}}
    \left[1-\cos\left(2\Theta t\right)\right]\,,
    \\
    {\cal F}(t)&=&\frac{\chi^{3}\theta}{2\Theta^{4}}
    \left[\cos\left(2\Theta t\right)-1\right]
    -\frac{\chi\theta}{\Theta^{4}}\left(\chi^{2}+\theta^{2}\right)
    \left[\cos\left(\Theta t\right)-1\right]
    +\overline{n}\frac{\chi\theta}{2\Theta^{2}}
    \left[1-\cos\left(2\Theta t\right)\right]\,.
\end{eqnarray}
\end{mathletters}

After an interaction time $t$, the state of the whole system 
can be expressed in terms of the normally ordered characteristic function as
\begin{equation}\label{eq:rho1b2}
    {\hat\rho}_{1b2}=\int\frac{d^{2}\mu}{\pi}
    \int\frac{d^{2}\nu}{\pi}\int\frac{d^{2}\zeta}{\pi}
    \Phi(\mu,\nu,\zeta,t)e^{-|\mu|^{2}-|\nu|^{2}-|\zeta|^{2}}
    {\hat D}_{1}(-\mu){\hat D}_{b}(-\nu){\hat D}_{2}(-\zeta)\,,
\end{equation}
where ${\hat D}$ indicates normally ordered displacement operator.

\section{Teleportation protocol}

The idea is to find an experimentally feasible, 
{\em modified} version of the standard protocol for the teleportation 
of continuous quantum variables \cite{VAI94,BRA98}, able to minimize 
the disturbing effects of the beam-splitter-type term in Eq.~(\ref{eq:Heff}).

First of all, the driving mode is filtered out
after reflection on the mirror
(see Fig.\ref{fig1}),
allowing only the modes ${\hat a}_{1}$ and ${\hat a}_{2}$
to reach Alice's station.
Then Alice performs a heterodyne measurement \cite{YUE80}
on the mode ${\hat a}_{2}$, projecting it onto a coherent
state $|\alpha\rangle$.
Alice and Bob are left with an entangled state
for the optical Stokes mode $a_{1}$ and the vibrational mode $b$,
conditioned to this measurement result, i.e.,
\begin{equation}\label{eq:rho1b}
    {\hat\rho}_{1b}=N\int\frac{d^{2}\mu}{\pi}
    \int\frac{d^{2}\nu}{\pi}\int\frac{d^{2}\zeta}{\pi}
    \Phi(\mu,\nu,\zeta,t)e^{-|\mu|^{2}-|\nu|^{2}-|\zeta|^{2}}
    {\hat D}_{1}(-\mu){\hat D}_{b}(-\nu)
    \langle\alpha|{\hat D}_{2}(-\zeta)|\alpha\rangle\,,
\end{equation}
and the normalization constant is 
$N=({\cal E}+1)\exp\left[|\alpha|^{2}/({\cal E}+1)\right]$.
Denoting with ${\tilde\Phi}(\mu,\nu)$ the normally ordered 
characteristic function associate to the state (\ref{eq:rho1b}),
we have 
\begin{eqnarray}\label{eq:Phitil}
    {\tilde\Phi}(\mu,\nu)&=&N\int\frac{d^{2}\zeta}{\pi}
    \Phi(\mu,\nu,\zeta) e^{-\zeta\alpha^{*}+\zeta^{*}\alpha-|\zeta|^{2}}
    \nonumber\\
    &=&\exp\left[-\left({\cal A}-\frac{{\cal F}^{2}}{{\cal E}+1}\right)
    |\mu|^{2}
    -\left({\cal B}-\frac{{\cal D}^{2}}{{\cal E}+1}\right)
    |\nu|^{2}\right.
    \nonumber\\
    &&\left.+\left({\cal C}+\frac{\cal FD}{{\cal E}+1}\right)
    \left(\mu\nu+\mu^{*}\nu^{*}\right)
    +\frac{\cal F}{{\cal E}+1}\left(\alpha\mu-\alpha^{*}\mu^{*}\right)
    +\frac{\cal D}{{\cal E}+1}\left(\alpha\nu^{*}-\alpha^{*}\nu\right)
    \right]\,.
\end{eqnarray}

Introducing the quadratures 
\begin{mathletters}\label{eq:qua}
\begin{eqnarray}
    {\hat X}_{a_{1}}&=&\frac{{\hat a}_{1}+{\hat a}_{1}^{\dag}}{\sqrt{2}}\,,
    \quad
    {\hat P}_{{\hat a}_{1}}=\frac{{\hat a}_{1}
    -{\hat a}_{1}^{\dag}}{i\sqrt{2}}\,,
    \\
    {\hat X}_{b}&=&\frac{{\hat b}+{\hat b}^{\dag}}{\sqrt{2}}\,,
    \quad
    {\hat P}_{b}=\frac{{\hat b}-{\hat b}^{\dag}}{i\sqrt{2}}\,.
\end{eqnarray}
\end{mathletters}
it is possible to evaluate their correlations through 
Eq.(\ref{eq:Phitil}). In particular,
defining ${\hat {\bf v}}=({\hat X}_{a_{1}},{\hat P}_{a_{1}},{\hat 
X}_{b},{\hat P}_{b})$,
the correlation matrix
$\Gamma_{i,j}=\langle{\hat{\bf v}}_{i}{\hat{\bf v}}_{j}
+{\hat{\bf v}}_{j}{\hat{\bf v}}_{i}\rangle/2$ results
\begin{equation}
    \Gamma=\left(
    \begin{array}{cccc}
    {\cal A}-\frac{{\cal F}^{2}}{{\cal E}+1}+\frac{1}{2}&0
    &{\cal C}+\frac{\cal FD}{{\cal E}+1}&0
    \\
    0&{\cal A}-\frac{{\cal F}^{2}}{{\cal E}+1}+\frac{1}{2}
    &0&-{\cal C}-\frac{\cal FD}{{\cal E}+1}
    \\
    {\cal C}+\frac{\cal FD}{{\cal E}+1}&0
    &{\cal B}-\frac{{\cal D}^{2}}{{\cal E}+1}+\frac{1}{2}&0
    \\
    0&-{\cal C}-\frac{\cal FD}{{\cal E}+1}
    &0&{\cal B}-\frac{{\cal D}^{2}}{{\cal E}+1}+\frac{1}{2}
    \end{array}
    \right)\,.
    \label{eq:Gam}
\end{equation}

We now employ the standard protocol for the teleportation of 
continuous quantum variables \cite{VAI94,BRA98}.
The quantum channel between Alice and Bob is established via
two-mode entangled state described by the correlation matrix
(\ref{eq:Gam}).

An input Gaussian state at Alice's side
can be fully described by its $2\times 2$
covariance matrix $\Gamma^{in}$.
Then, the output Gaussain state at Bob's side would be characterized 
by the covariance matrix $\Gamma^{out}$.
The input-output relation for these matrices can be found as
follows. In terms of normally ordered characteristic functions we have
\begin{equation}
    \exp\left[-\frac{1}{4}{\bf u}\Gamma^{out}{\bf u}^{T}\right]
    =\tilde{K}({\bf u})
    \exp\left[-\frac{1}{4}{\bf u}\Gamma^{in}{\bf u}^{T}\right]
\end{equation}
where ${\bf u}=(q,p)$ is the variable vector of the characteristic 
functions. Instead $\tilde{K}$ is the Fourier transform of the kernel
in the integral transform mapping the
Wigner function of the input state into the
Wigner function of the output state (see e.g. Ref.\cite{CHI02}).
In terms of the Wigner function $W_{AB}$ of the state shared by 
Alice and Bob, it results
\begin{eqnarray}
    \tilde{K}({\bf u})&=&
    \int dx_{A} dp_{A} dx_{B} dp_{B}\;
    e^{-ix_{A}q-ix_{B}q+ip_{A}p-ip_{B}p}\;
    W_{AB}(x_{A},p_{A},x_{B},p_{B})
    \nonumber\\
    &=&\exp\left[-\frac{1}{4}\left(q,-p,q,p\right)
    \Gamma\left(q,-p,q,p\right)^{T}\right]\,.
\end{eqnarray}
Then, it is easy to derive the relations
\begin{mathletters}\label{eq:Gout}
\begin{eqnarray}
    \Gamma_{11}^{out}&=&\Gamma_{11}^{in}+\left(
    \Gamma_{11}+2\Gamma_{13}+\Gamma_{33}\right)\,,
    \\
    \Gamma_{12}^{out}&=&\Gamma_{12}^{in}+\left(
    \Gamma_{14}-\Gamma_{12}+\Gamma_{34}-\Gamma_{23}\right)\,,
    \\
    \Gamma_{22}^{out}&=&\Gamma_{22}^{in}+\left(
    \Gamma_{22}-2\Gamma_{24}+\Gamma_{44}\right)\,.
\end{eqnarray}
\end{mathletters}
Thus, the fidelity of the teleportation protocol can be written, with the 
help of Eqs.(\ref{eq:Gam}) and (\ref{eq:Gout}) as 
\begin{equation}
    F=\frac{1}{1+\left[1+{\cal A}(t)+{\cal B}(t)+2{\cal C}(t)
    -({\cal F}-{\cal D})^{2}/({\cal E}+1)\right]}\,,
    \label{eq:Fid}
\end{equation}
where we have specialized to the case of an input coherent state.
In such a case, the upper bound for the fidelity achievable 
with only classical means and no quantum resources is $F=1/2$
\cite{BRA99}.

The fidelity (\ref{eq:Fid}) does not depend on the Bob's local 
operations. In fact these are merely displacements based on the 
Alice's  measurement results $X_{+},P_{-},\alpha$, i.e.
${\hat X}_{b}\to {\hat X}_{b}+\sqrt{2}X_{+}
+\sqrt{2}{\rm Re}\{\alpha\}({\cal F}-{\cal D})/({\cal E}+1)$,
${\hat P}_{b}\to {\hat P}_{b}-\sqrt{2} P_{-}
+\sqrt{2}{\rm Im}\{\alpha\}({\cal F}+{\cal D})/({\cal E}+1)$.
Note that the amount proportional to ${\cal F}/({\cal E}+1)$
deserves to account for the shifted results $X_{+},P_{-}$
obtained by Alice by virtue of the heterodyne detection
(see Eq.~(\ref{eq:Phitil})), 
while the amount proportional to ${\cal D}/({\cal E}+1)$
deserves to cancel the diplacement on the ${\hat b}$ 
mode caused again by the heterodyne detection 
(see Eq.~(\ref{eq:Phitil})).

To actuate the phase-space displacement, 
Bob can use again the radiation pressure force.
In fact, if the mirror is shined by a bichromatic intense laser
field with frequencies $\varpi_{0}$ and
$\varpi_{0}+\Omega$, employing again Eq.~(\ref{eq:Hini})
and the RWA, one is left with an effective interaction Hamiltonian
\begin{equation}
H_{act} \propto {\hat b}e^{-i\varphi}+{\hat b}^{\dag}e^{i\varphi},
\end{equation}
where $\varphi$ is the relative phase between the two frequency 
components. Any phase space displacement of the mirror vibrational 
mode can be realized by adjusting this relative phase and the 
intensity of the laser beam.

Finally, for what concerns the experimental verification of
teleportation, that is, the measurement of the final state of the 
acoustic mode,
one can consider a second, intense ``reading'' laser pulse,
and exploit again the optomechanical interaction given by
Eq.~(\ref{eq:Heff}), where now $a_{1}$ and $a_{2}$ are meter modes.
It is in fact possible to perform a heterodyne measurement 
\cite{YUE80} of an
appropriate combination of the two back-scattered modes,
${\hat Z}={\hat a}_{1}-{\hat a}_{2}^{\dag}$, if
the driving laser beam at frequency $\omega_0$ is used as local
oscillator and the resulting photocurrent is mixed with a signal oscillating
at the frequency $\Omega$.
The behaviour of $Z(t)$ as a function of the time duration of the
second ``measuring'' driving beam can be derived from Eqs.~(\ref{eq:sol}), 
that is
\begin{eqnarray}
{\hat Z}(t)\equiv {\hat a}_{1}(t)-{\hat a}_{2}^{\dag}(t)&=&
\frac{1}{\Theta}\left[\chi+\theta\right]\sin(\Theta t) 
{\hat b}^{\dag}(0)
\nonumber\\
&+&\frac{1}{\Theta^{2}}\left[
\theta^{2}-\chi^{2}\cos(\Theta t)-\chi\theta+\chi\theta\cos(\Theta 
t)\right]{\hat a}_{1}(0)
\nonumber\\
&-&\frac{1}{\Theta^{2}}\left[
\chi\theta+\chi\theta\cos(\Theta t)-\chi^{2}-\theta^{2}\cos(\Theta 
t)\right]{\hat a}_{2}^{\dag}(0)\,.
\end{eqnarray}
It is easy to see that for $\cos(\Theta t)=0$ and 
$\Theta(\theta+\chi)\gg\theta(\theta-\chi)$
the measured quantity practically coincides with the mode 
oscillation operator $b^{\dag}(0)$, thus revealing information on the 
state of the mechanical oscillator.

\section{Results and Conclusions}

\begin{figure}[t]
\centerline{\epsfig{figure=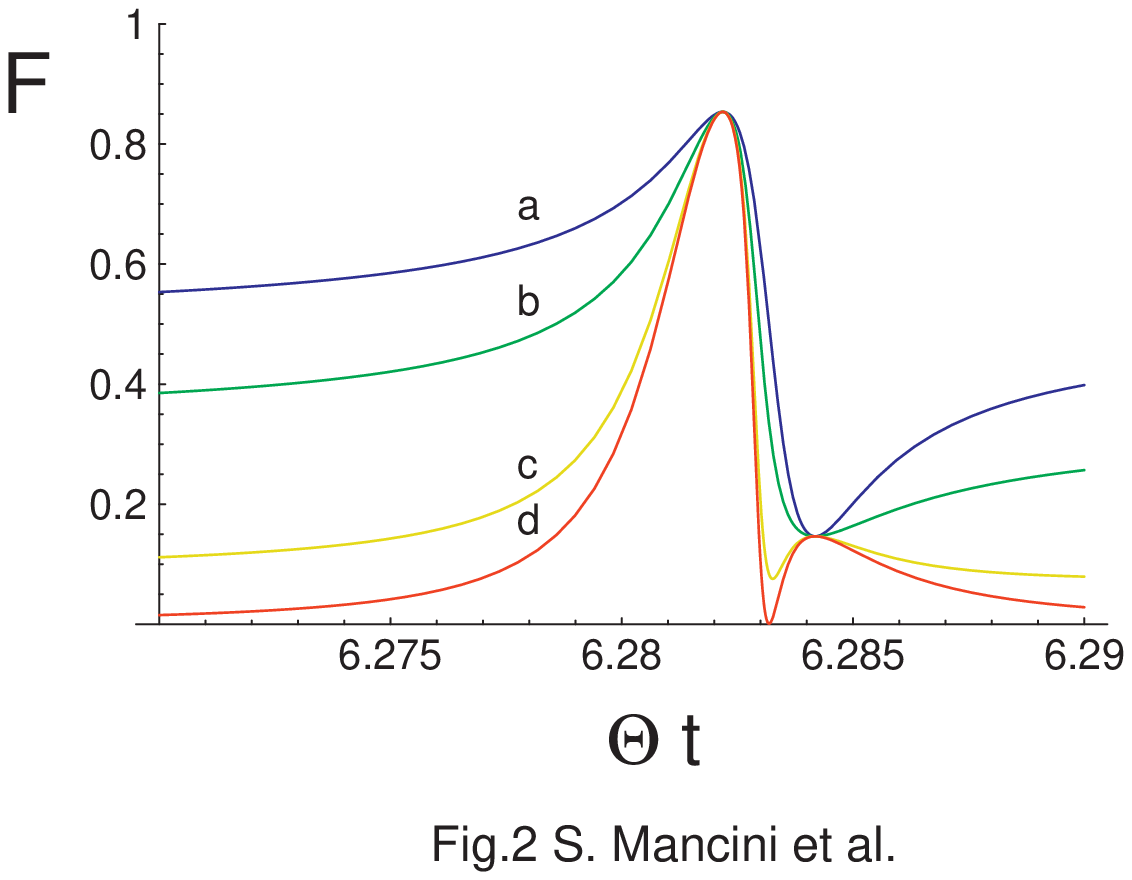,width=3.0in}}
\caption{\widetext
Fidelity $F$ vs the scaled time $\Theta t$.
Curves a, b, c, d are for $\overline{n}=0$,
1, 10, $10^{3}$, respectively.
The values of parameters are: $\wp=10$ W;
$\Omega=5\times 10^{8}$ Hz;
$\Delta\nu_{det}=10^{7}$ Hz; $M=10^{-10}$ Kg;
$\omega_{0}= 2\times 10^{15}$ Hz, $\Delta\nu_{mode}=10^{3}$ Hz.}
\label{fig2}
\end{figure}

Fig.~\ref{fig2}
shows the fidelity (\ref{eq:Fid}) as a function 
of the (rescaled) interaction time
$t$ for different values of the initial mean thermal phonon number of 
the mirror acoustic mode $\overline{n}$.
The fidelity $F$ is periodic
in the interaction time $t$ (see Methods),
and we show only one of all possible time windows where $F$ reaches
its maximum. The remarkable result shown in Fig.~\ref{fig2} 
is that this maximum value, 
$F_{max} \simeq 0.85$, is well above the classical bound $F=0.5$ and that it 
is surprisingly independent of the initial temperature of the acoustic mode.
This is apparently in contrast with previous results \cite{DUA00}
showing that entanglement is no longer useful above
one thermal photon (or phonon).
This effect could be ascribed to quantum interference
phenomena, and opens the way 
for the demonstration of quantum teleportation of states
of macroscopic systems.
However, thermal noise has still important effects so that, in 
practice, any experimental implementation needs an 
acoustic mode cooled at low temperatures (see however 
Refs.~\cite{VIT02,COH99}
for effective cooling mechanism of acoustic modes).
In fact, we see from Fig.~\ref{fig2} that by increasing $\overline{n}$,
the useful time interval becomes narrower.
That means the necessity of designing precise driving laser pulses
in order to have a well defined interaction time.
Furthermore, the time interval within
which the classical communication from Alice to Bob, and the phase 
space displacement by Bob have to be made, becomes shorter and shorter
with increasing temperature, because the vibrational state projected
by Alice's Bell measurement heats up in a time of the order 
of $(\gamma_{m}\overline{n})^{-1}$,
where $\gamma_{m}$ is the mechanical damping constant. The effects of 
mechanical damping can be instead neglected during the 
back-scattering process stimulated by the intense laser beam. In fact, 
mechanical damping rates of about $\gamma_{m} \simeq 1$ Hz are
available, and therefore negligible with respect to
the typical values of the coupling constants 
$\chi \simeq \theta \simeq 5\times 10^{5}$ Hz, and $\Theta \simeq 
10^{3}$ Hz, determining the Hamiltonian dynamics (see Methods).
Such values are obtained with the following choice of parameters:
${\wp} = 10$W, $\omega_{0}\sim 2\times 10^{15}$ Hz, $\Omega \sim 
5\times 10^{8}$ Hz, $\Delta\nu_{det}\sim 10^{7}$ Hz, 
$\Delta\nu_{mode}\sim t^{-1} \sim 10^{3}$ Hz, and $M \sim 10^{-10}$ 
Kg, which are those used in Fig.~\ref{fig2}. These parameters
are slightly different from those of already 
performed optomechanical experiments \cite{TIT99,COH99}.
However, using a thinner silica crystal and considering higher
frequency modes, the parameters we choose could be obtained. These choices
show the difficulties one meets in trying to extend
genuine quantum effects as teleportation into the macroscopic domain.

The continuous variable teleportation protocol presented here modifies 
the standard one of Refs.~\cite{VAI94,BRA98} by adding a heterodyne 
measurement on the ``spectator'' mode ${\hat a}_{2}$. This additional 
measurement performed by Alice is important because it significantly 
improves the teleportation protocol. In fact, it is easy to see that 
if no measurement is performed on the anti-Stokes mode, 
the resulting fidelity for the teleportation of coherent states 
is always smaller with respect to that with the heterodyne measurement.
In particular, there is still a maximum value of the fidelity,
$F_{max}=0.80$ in this case, independent of temperature, 
but the useful interaction time interval becomes much narrower for 
increasing temperature. 

It is worth remarking that the present teleportation scheme provides
also a very powerful {\em cooling} mechanism for the acoustic mode. 
As matter of fact, its effective number of thermal excitations soon after 
the two homodyne measurements at Alice station becomes
$\overline{n}_{eff}=1+{\cal A}+{\cal B}+2{\cal C}
-({\cal F}-{\cal D})^{2}/({\cal E}+1)$.
It reduces to $\overline{n}+1$ in absence of entanglement,
where $1$ represents the noise introduced by the protocol.
Instead, the optomechanical interaction for a proper time permits
to achieve $\overline{n}_{eff}=0.17$, i.e., an $80\%$ 
reduction of thermal noise
at once, at the moment of Alice's measurement, thanks to the 
entanglement. 
To this end, the classical communication and the
phase space displacement at Bob's site are 
unnecessary, since they do not affect the state variances.

In conclusion, we have proposed a simple scheme to teleport an
unknown quantum state of a radiation field
onto a macroscopic, collective vibrational 
degree of freedom of a massive mirror.
The basic resource of
entanglement is attained by means of the optomechanical coupling 
provided by the radiation pressure. 
Here we have shown the 
teleportation of the quantum information contained in an unknown 
quantum state of a radiation field to a collective degree of freedom 
of a massive object. This scheme could be easily extended in principle 
to realize a transfer of quantum information between two massive 
objects. In fact
Victor could use tomographic reconstruction schemes, 
again based on the ponderomotive interaction (see \cite{MAN96}), to 
``read'' the quantum state of a vibrational mode of another mirror
and use this information to prepare the state of the radiation field 
to be sent to Alice.
The present result could be challenging tested
with present technology, and opens new perspectives towards the use
of quantum mechanics in macroscopic world.
For example, we recognize possible
technological applications such as the
preparation of nonclassical states of
micro-electro-mechanical systems (MEMS) \cite{CLE98},
where the oscillation frequency could be higher and, consequently, the
working temperature can be raised.

\end{document}